%% file: paper.tex
\documentclass[submission,copyright,creativecommons]{eptcs}
\providecommand{\event}{LMFTP 2021} 
\usepackage{underscore}           
\usepackage[misc]{ifsym}
\usepackage[utf8]{inputenc}
\usepackage[T1]{fontenc}
\usepackage{xspace}
\usepackage{enumitem}
\usepackage{colonequals}
\usepackage{stmaryrd}
\usepackage{todonotes}
\usepackage{amsmath,amscd,amsbsy,amssymb,latexsym,url,bm,amsthm}
\usepackage{epsfig,graphicx,subfigure}
\usepackage{hyperref}
\usepackage{xcolor}
\usepackage{appendix}
\usepackage[vlined,ruled,commentsnumbered,linesnumbered]{algorithm2e}
\usepackage{enumerate,proof}

\usepackage{listings}

\lstdefinelanguage{Coq}%
  {morekeywords={Variable,Inductive,CoInductive,Fixpoint,CoFixpoint,%
      Definition,Lemma,Theorem,Axiom,Goal,Local,Save,Grammar,Syntax,Intro,%
      Trivial,Qed,Intros,Symmetry,Simpl,Rewrite,Apply,Elim,Assumption,%
      Left,Cut,Case,Auto,Unfold,Exact,Right,Hypothesis,Pattern,Destruct,%
      Constructor,Defined,Fix,Record,Proof,Induction,Hints,Exists,let,in,%
      Parameter,Split,Red,Reflexivity,Transitivity,if,then,else,Opaque,%
      Transparent,Inversion,Absurd,Generalize,Mutual,Cases,of,end,Analyze,%
      AutoRewrite,Functional,Scheme,params,Refine,using,Discriminate,Try,%
      Require,Load,Import,Scope,Open,Section,End,match,with,Ltac,%
      Instance,Class,%
      bind,as,%
	},%
   sensitive, %
   morecomment=[n]{(*}{*)},%
   morestring=[d]",%
   literate={=>}{{$\Rightarrow$}}1 {>->}{{$\rightarrowtail$}}2{->}{{$\rightarrow$}}1
   {forallx}{{$\forall\!\!$}}1
  }[keywords,comments,strings]%

\title{Countability of Inductive Types Formalized \\ in the Object-Logic Level}
\author{Qinxiang Cao \and Xiwei Wu \thanks{Parallel authorship, equal contribution. Corresponding author: Qinxiang Cao.}}
\def\titlerunning{Countability}
\def\authorrunning{Qinxiang Cao \& Xiwei Wu}

\begin{document}
\maketitle

\begin{abstract}
  \textbf{Abstract:}  The set of integer number lists with finite length, and the set of binary trees with integer labels are both countably infinite. Many inductively defined types also have countably many elements. In this paper, we formalize the syntax of first order inductive definitions in Coq and prove them countable, under some side conditions. Instead of writing a proof generator in a meta language, we develop an axiom-free proof in the Coq object logic. In other words, our proof is a dependently typed Coq function from the syntax of the inductive definition to the countability of the type. Based on this proof, we provide a Coq tactic to automatically prove the countability of concrete inductive types. We also developed Coq libraries for countability and for the syntax of inductive definitions, which have value on their own.

\end{abstract}

\input{intro}
\input{premilinary}

\input{formalize}
\input{automation}
\input{relatedwork}
\input{conclusion}
\paragraph{Ackownledgement} This research is sponsored by National Natural Science foundation of China (NSFC) Grant No. 61902240 and Shanghai Pujiang Program 19PJ406000.

\nocite{*}
\bibliographystyle{eptcs}
\bibliography{fulllist}

\end{document}

%% file: intro.tex
\section{Introduction} \label{sec:intro}
In type theory, a system supports inductive types if it allows users to define new types from constants and functions that create terms of objects of that type.
The Calculus of Inductive Constructions (CIC) is a powerful language that aims to represent both functional programs in the style of the ML language and proofs in higher-order logic \cite{paulin2015introduction}.
Its extensions are used as the kernel language of Coq \cite{barras1999coq} and Lean \cite{DBLP:conf/cade/MouraKADR15}, both of which are widely used, and are widely considered to be a great success.
In this paper, we focus on a common property, countability, of all first-order inductive types, and provide a general proof in Coq's object-logic. 
The techniques that we use in this formalization can be useful for formally proving other properties of inductive types in the future.

Here we show some examples of inductive types that we will use in this paper :
\begin{lstlisting}
    Inductive natlist :=  Cons : nat -> natlist -> natlist
                          | Nil  : natlist.
    Inductive bintree :=  Node : nat -> bintree -> bintree -> bintree
                          | Leaf : bintree.
    Inductive expr := andp   : expr -> expr -> expr
                      | orp    : expr -> expr -> expr
                      | impp   : expr -> expr -> expr
                      | falsep : expr 
                      | varp   : nat -> expr.
\end{lstlisting}
As demonstrated above, \lstinline{natlist} (list of natural numbers), \lstinline{bintree} (binary trees with natural numbers as labels) and \lstinline{expr} (the expressions of propositional language with \lstinline{nat} as variable identifiers) can be defined inductively in the Coq proof assistant \cite{barras1999coq}.
Here, the word ``inductive'' means that \lstinline{natlist}, \lstinline{bintree} and \lstinline{expr} are the smallest of all sets that satisfy the above typing constraints. Specifically, \lstinline{natlist} is defined as the smallest set containing ``\lstinline{Nil}'' and closed by the ``\lstinline{Cons}'' constructor, which allows us to define a recursive function \lstinline{length} on \lstinline{natlist} so that \lstinline{length Nil = 0}, \linebreak\lstinline{length (Cons x l) = 1 + length l}, and to prove related properties by induction on the structure of \lstinline{natlist}. 

Countability is a basic property of sets. To state that a set is countable means that it has the same cardinality as some subset of the set of natural numbers.
For example, in the Henkin style proof of FOL completeness \cite{ebbinghaus2013mathematical}, one important step is to construct a maximal consistent set $\Psi$ by expanding a consistent set $\Phi$ of propositions:
\begin{eqnarray*}
\Psi_0 &:=& \Phi \\
\Psi_{n+1} &:=&\Psi_n \cup \{\phi_n\} \quad (\text{if } \Psi_n \cup \phi_n \text{ is consistent}) \\
\Psi_{n+1} &:=&\Psi_n \quad \ \ \ \ \ \ \ \ \ \ (\text{if } \Psi_n \cup \phi_n \text{ is inconsistent}) \\
\Psi & := & \bigcup_n \Psi_n 
\ \ \ \ \ \ \ \ \ \text{where } \phi_0, \phi_1, \dots \text{ are all FOL propositions.}
\end{eqnarray*}

In that proof, it is critical that the set of all FOL propositions is countable. The countability property allows us to enumerate all propositions as $\phi_0, \phi_1, \dots$.
As another example, 
we can show the countability of computable functions by proving the countability of untyped lambda expressions.
Since the set of all functions from natural numbers to natural numbers is uncountable, there must exist at least one uncomputable function.
Like FO-propositions and lambda expressions, many sets can be formalized as inductive types, and we focus on the countability of the inductive types in this paper.

Proving \lstinline{natlist} to be countable is straightforward.
(1) The only  \lstinline{natlist} of length $0$ is  \lstinline{Nil}.
(2) The \lstinline{natlist}s of length $(n+1)$ should be countable if those \lstinline{natlist}s of length $n$ are countable (because the former set is isomorphic with the Cartesian product of $\mathbb{N}$ and the latter set).
(3) By induction, the set of  \lstinline{natlist}s of length $n$ is countable for any $n$.
(4) The set of all \lstinline{natlist}s is a union of countably many countable sets, and thus is countable (because we can easily construct an bijection from $\mathbb{N}^2$ to $\mathbb{N}$: $f(x,y) = 2^x (2y + 1) - 1$ and the construction does not even need the choice axiom).
Similarly, natural number labeled binary trees with size $n$ are countable for any $n$. Thus, the elements of \lstinline{bintree} are countable.
It is natural to apply the same proof idea to a more complex inductive type. We define a \emph{rank} function to generalize the length function for \lstinline{natlist} and the size function for binary trees (with some slight modification). For example, the \lstinline{rank} function on \lstinline{natlist} and \lstinline{bintree} satisfies:
\begin{lstlisting}
    rank Nil = 1
    rank (Cons n l) = rank l + 1
    rank Leaf = 1
    rank (Node n l r) = rank l + rank r + 1
\end{lstlisting}
We prove that given a fixed inductive type, its elements with rank less than $n$ are countable\footnote{In the general proof, we consider elements with rank less than $n$, not elements with rank equal to $n$}. Then, all elements of this type are also countable since the set is a union of countably many countable sets. One could write Coq tactics, which is a meta language, to describe our proof ideas above. In contrast, our target in this paper is to formally prove one single theorem in the object language for general inductive types' countability.

Handling general inductive types in Coq's object language is hard.
Coq's object language, Gallina, has built-in support for recursive functions and inductive proofs, as long as they are about concrete inductive types.
For general inductive types, we do not have such support, and even simple pattern match expressions are not easy to formalize.
We choose to derive recursive functions and inductive proof principles from general recursive functions.
Using concrete inductive types as an example, \lstinline{natlist}'s general recursive function is:
\begin{lstlisting}
    natlist_rect
       : forall P : natlist -> Type,  P Nil ->
         (forall (n : nat) (l : natlist), P l -> P (Cons n l)) ->
         forall l : natlist, P l
\end{lstlisting}
It satisfies, for any (maybe dependently typed) $P$ and $F_0$, $F_1$,
\begin{lstlisting}[mathescape]
    natlist_rect $P$ $F_0$ $F_1$ Nil = F_0
    natlist_rect $P$ $F_0$ $F_1$ (Cons n l) = F_1 n l (natlist_rect $P$ $F_0$ $F_1$ l)
\end{lstlisting}
We generalize the combination of \lstinline{natlist_rect} and the two equalities above, and develop our proofs based on them.

Theoretically, it is more difficult to do something at the object-logic level than at the meta-logic level.
Any proof formalized at the object-logic level is a Coq function from its assumptions to its conclusion, according to Curry-Howard correspondence.
One can always develop a corresponding meta-language function that implements the ``same'' functionality.
In contrast, some statements are only provable in a meta-logic, but are unprovable in the object logic.
Martin Hofmann and Thomas Streicher showed that the principle of uniqueness of identity proofs is not derivable in the object logic itself \cite{316071}. 

Practically, our proof automation, which uses object-logic proofs, is more efficient than proof generators written in a meta-language.
Our tool can prove \lstinline{expr} countable in Coq in 0.089 seconds but a tactic-based proof will take 1.928 seconds to finish the proof\footnote{Processor: 2.3 GHz 8-Core Intel Core i9; Memory: 16 GB 2400 MHz DDR4.} (see Coq development for more details).
This result arises because our tool only requires Coq to typecheck one theorem with its arguments, but proofs, either proof scripts or proof terms, generated by a meta-language generator require Coq to typecheck every single proof step.

In this paper, we formalize the general proof of countability theorem mentioned above, using Coq, and automate our proof to avoid repeating the long proof process.
There are several ways in which this goal may be achieved. 
One is to use external tools to generate the operations and proofs of corresponding lemmas.
For example, DBGen \cite{polonowski2013automatically} generates single-variable substitution operations, and 
Autosubst2 \cite{stark2019autosubst} can generate substitution-related definitions and Coq proof terms.
Another approach is to use the internal facility of theorem provers, which is written in a built-in meta language, to generate proof terms or proof scripts.
Brian Huffman and Alexander Krauss (old datatype), and Jasmin Blanchette
(BNF datatype) have developed tactics, which is a meta language, to prove
datatypes countable \cite{0Deriving} in Isabelle/HOL. 

In comparison, an object logic proof of ``for any possible $T$, $P(T)$ holds'' is one singleton proof term of type $\forall T, P(T)$.
A meta-logic proof is a meta-level program (with probably more expressiveness power) which takes $T$ as its input and outputs a proof term of $P(T)$, which could be huge.
Intuitively, the former one directly \emph{states} that $\forall T, P(T)$ is true, while the latter one is an oracle which can \emph{step-by-step explain} why $P(T)$ holds for a concrete $T$.
Arthur Azevedo de Amorim's implementation \cite{coqPL2020} is the only object-level proof of countability before our paper.
He used indices to number constructors of an inductive type. In other words, he formalized the syntax of inductive types and deeply embedded the syntax (in some sense, the meta language) in Coq's object language.
As a consequence, his proof involves complicated reasoning about indices' equivalence, type's equivalence and dependent type issues---if two Coq types $T_1 = T_2$, $T_1$'s elements are not automatically recognized as $T_2$'s elements by Coq's type checker.
Our formalization shows that such proof-reflection technique is not needed for a general countability proof.

\noindent \textbf{Contributions.} Our main contributions are a Coq formalized general countability proof for first order inductive types, and an automatic tactic for proving inductive types countable. We do not need any external tool to generate definitions, proof terms, or proof scripts, and our proof itself does not involve complicated dependently typed reasoning about type equalities. We also developed Coq libraries for countability and for a syntax of inductive definitions, which have values of their own. All of our proofs are formalized axiom-free, and our proof of countability theorem can be used in the completeness proof of separation logics, a Coq formalization for an early paper \cite{cao2017bringing}.

\noindent \textbf{Outline.} In Section \ref{sec:pre}, we will clarify our Coq definition of countability and our formalization of the syntax of first order inductive type definitions. 
In Section \ref{sec:form}, we present our general countability theorem and our proof.
In Section \ref{sec:autom}, we introduce our automatic tactic for proving concrete inductive types countable.
We discuss related works in Section \ref{sec:related} and conclude in Section \ref{sec:concl}.

%% file: premilinary.tex
\section{Preliminaries} \label{sec:pre}
In this section, we present our formal definition of countable (Section \ref{subsec:count}),  the syntax of  first order inductive types (Section \ref{subsec:syntax}), and general recursive functions (Section \ref{subsec:rect}).
We will also list their important properties, that we prove in our Coq library.

\subsection{Countable} \label{subsec:count}
We define the type T to be countable if and only if there exists an injection from T to natural numbers, which means T is either finite or countably infinite. Here, an injection is a relation that keeps the injective property and functional property. This \lstinline{Countable} is the definition used in our final theorem, but we use an auxiliary definition, \lstinline{SetoidCountable}, in our proof.
For the countability proof of inductive type $T$, we need to prove that $\{x: T \mid \mathsf{rank}(x) < n\}$ is countable for any $n$.
In Coq, an element in $\{x: T \mid \mathsf{rank}(x) < n\}$ is a dependently typed tuple $(x, p)$, where $x \in T$ and $p$ is a proof of $\mathsf{rank}(x) < n$.
Two such dependently typed tuples $(x_1, p_1)$ and $(x_2, p_2)$ are equal if $x_1 = x_2$, and $p_1$ and $p_2$ are identical proof terms.
Proving $\{x: T \mid \mathsf{rank}(x) < n\}$ \lstinline{Countable} requires us to show whether two proofs, $p_1$ and $p_2$, of $\mathsf{rank}(x) < n$ are identical.
Using \lstinline{SetoidCountable} avoids that kind of reasoning about proof terms, and avoids using the ``proof-irrelevance'' axiom in some sense\footnote{\lstinline{Axiom proof_irrelevance : forall (P:Prop) (p1 p2:P), p1 = p2.}}.
The definition of $\{x: T \mid \mathsf{rank}(x) < n\}$ being \lstinline{SetoidCountable} is straightforward: there exists a function $f$ from $\{x: T \mid \mathsf{rank}(x) < n\}$ to natural numbers, so that if $f(x_1, p_1) = f(x_2, p_2)$ then $x1 = x2$.

\begin{lstlisting}
    Definition image_defined {A B} (R: A -> B -> Prop): Prop :=
        forall a, exists b, R a b.
    Definition partial_functional {A B} (R: A -> B -> Prop): Prop :=
        forall a b1 b2, R a b1 -> R a b2 -> b1 = b2.
    Definition injective {A B} (R: A -> B -> Prop): Prop :=
        forall a1 a2 b, R a1 b -> R a2 b -> a1 = a2.
    Record injection (A B: Type): Type := {
        inj_R:> A -> B -> Prop;
        im_inj: image_defined inj_R;
        pf_inj: partial_functional inj_R;
        in_inj: injective inj_R }.
    Definition Countable (T : Type) := injection T nat.
    Record Setoid_injection
      (A B: Type) (RA: A -> A -> Prop) (RB: B -> B -> Prop) := ...
        (* RA and RB are equivalence relations on A and B resp. *)
    Definition SetoidCountable (A: Type) {RA: A -> A -> Prop}: Type :=
      @Setoid_injection A nat RA (@eq nat).
        (* RA is an equivalence relation on A *)
        (* @SetoidCountable A RA if the quotient set A/RA is countable. *)
\end{lstlisting}

In our countability library, we prove that products of two countable types and unions of countably many countable types are countable. 
We prove that the composition of two injections is still an injection (see \lstinline{injective_compose} below), and if $f \circ g$ is an injection then $f$ is an injection (see \lstinline{injective_compose_rev} below). We also prove their setoid versions, but omit them here.
We define ``bijection'' and prove some elementary properties about bijection and injection.
For connections between \lstinline{SetoidCountable} and \lstinline{Countable}, we prove that any \lstinline{Setoid_injection} on Coq's builtin equality is an \lstinline{injection}, and thus any \lstinline{SetoidCountable} type w.r.t. Coq's builtin equality is also \lstinline{Countable}.

\begin{lstlisting}
    Lemma injective_compose {A B C} (R1: A -> B -> Prop) (R2: B -> C -> Prop):
        injective R1 -> injective R2 -> injective (compose R1 R2).
    Lemma injective_compose_rev {A B C} (R1: A -> B -> Prop) (R2: B -> C -> Prop):
        image_defined R2 -> injective (compose R1 R2) -> injective R1.
    Lemma SetoidCountable_Countable {A: Type}:
        SetoidCountable A (@eq A) -> Countable A.
\end{lstlisting}

Here, we use a relation rather than a function to provide the definition of \lstinline{Countable}, for better usability and extensibility. For example, if we have a bijection from $A$ to $B$ and we have \lstinline{Countable B}, we want to show that $A$ is also countable. We can do it easily by relation, but we cannot do it under the definition of function, because of the problem of computability.

\subsection{Syntax of inductive definition} \label{subsec:syntax}
In our formalization, we only consider \emph{first order inductive definitions}.
Not all inductive types have only countably many elements. Thus we exclude definitions like the following:
\begin{lstlisting}
    Inductive inf_tree: Type :=
    | inf_tree_leaf: inf_tree
    | inf_tree_node: nat -> (nat -> inf_tree) -> inf_tree.
\end{lstlisting}
However, we try to focus on techniques of building dependently type functions from inductive definitions to nontrivial proof terms in this work. Thus we choose to exclude mutually inductive definitions and nested inductive definitions from our formalization, although we believe that we can extend our work in the future to handle these cases.

Formally, a first-order inductive type $T$ is defined by a list of constructors, each of which is a first-order function with result type $T$.
The argument types of constructors should be constant base type\footnote{Here, constant base type means other types which are countable.} or $T$ itself.
So we formalize the syntax of an inductive definition $T$ as a list of dependently typed pairs of typing rules and constructors: \lstinline{list (sigT (fun arg => constr_type arg T))}, we will call it \lstinline{Constrs_type} later in this paper.
We usually call the ``typing rule'' part \lstinline{arg}, with a type \lstinline{list (option Type)}, and call a constructor \lstinline{constr}, whose type depends on \lstinline{arg} and is calculated by \lstinline{constr_type}.
For example, the definition of \lstinline{natlist} has two branches. The type of \lstinline{Cons} is:\linebreak \lstinline{nat -> natlist -> natlist}. It has two arguments, one of which is of type \lstinline{nat} and the other is the inductive type \lstinline{natlist} itself. Thus, this typing rule can be formalized as: \lstinline{[Some nat; None]} (\lstinline{Some} for a based type and \lstinline{None} for the inductive type itself) and
\begin{lstlisting}
  constr_type [Some nat; None] natlist = nat -> natlist -> natlist
\end{lstlisting}
exactly describes the type of constructor \lstinline{Cons}.
Since the definition of \lstinline{natlist} has two branches, this definition can be described by: \lstinline{[ _[ [Some nat; None], Cons ]_; _[ [], Nil ]_ ]}.

\subsection{General recursion} \label{subsec:rect}

For an inductive type \lstinline{T}, Coq generates \lstinline{T_rect}, \lstinline{T_ind}, \lstinline{T_rec} and \lstinline{T_sind}, which respectively correspond to elimination principles on \lstinline{Type}, \lstinline{Prop}, \lstinline{Set} and \lstinline{SProp}\cite{barras1999coq}. 
For our countability proof, \lstinline{T_rect} is enough\footnote{
  The constant \lstinline{T_ind} is always generated, whereas \lstinline{T_rec} and \lstinline{T_rect} may be impossible to derive, for example, when the sort is \lstinline{Prop}. However, we only focus on the inductive type so no problems are encountered.}.
We define \lstinline{rect_type} to compute the type of \lstinline{T_rect} (we call it \lstinline{rect} later in this paper) from an inductive definition. Similar to the definition of \lstinline{Constrs_type}, we use \lstinline{rect_clause_type} to compute each branch. Here we show the definitions of \lstinline{rect_type} :
\begin{lstlisting}         
  Fixpoint rect_type
         (T: Type) (constrs: Constrs_type) (P: T -> Type): Type :=
    match constrs with
    | nil => forall x: T, P x
    | _[ arg, constr ]_ :: constrs0 => 
         rect_clause_type arg T P constr -> rect_type T constrs0 P
    end.
\end{lstlisting}

For example, \lstinline{natlist_rect} (which is generated by Coq) has type
\begin{lstlisting}         
  forall P, rect_type natlist [ _[ [Some nat; None], Cons ]_; _[ [], Nil ]_ ] P
\end{lstlisting}

Knowledge of the type of general recursive function alone is not sufficient for building proofs.
It is important that the computation result of a recursive function coincides with the definitions in its corresponding branch.
For example, as mentioned in Section \ref{sec:intro}, \lstinline{natlist_rect} satisfies:
\begin{lstlisting}[mathescape]
    natlist_rect $P$ $F_0$ $F_1$ Nil = F_0
    natlist_rect $P$ $F_0$ $F_1$ (Cons n l) = F_1 n l (natlist_rect $P$ $F_0$ $F_1$ l)
\end{lstlisting}
For general inductive types, we need to consider all possible ways of filling \lstinline{rect}'s arguments.
We introduce \lstinline{apply_rect} so that \lstinline{apply_rect T P constrs para (rect P) x} fills \lstinline{rect}'s argument (like $F_0$ and $F_1$ above) in a parameterized way defined by \lstinline{para} and calculates the result on $x: T$.
Based on that, we define \lstinline{rect_correct}:
\begin{lstlisting}
    rect_correct (T: Type) (constrs: Constrs_type)
      (rect: forall P, rect_type T constrs P): Prop
\end{lstlisting}
to be the following property: for any \lstinline{para} and \lstinline{x}, \lstinline{apply_rect T P constrs para (rect P) x} equals to the recursive branch defined by \lstinline{para} and \lstinline{x} (and the recursive function \lstinline{apply_rect T P constrs} \lstinline{para rect} itself). Detailed definitions of \lstinline{apply_rect} and \lstinline{rect_correct} involve complicated dependent type issue, and we defer them to Section \ref{sec:form}.

In summary, our proof about inductive type's countability depends on and only depends on the following arguments and hypothesis:
\begin{itemize}
\item the Coq type \lstinline{T: Type};
\item the inductive definition \lstinline{constrs: Constrs_type};
\item the general recursive function \lstinline{rect: forall P, rect_type T constrs P};
\item the characteristic equations \lstinline{rect_correctness: rect_correct constrs rect}.
\end{itemize}
We develop three automatic tactics \lstinline{gen_constrs} , \lstinline{gen_rect} , \lstinline{apply_rect_correctness_gen} to get \lstinline{constrs}, \lstinline{rect} and \lstinline{rect_correctness} above from \lstinline{T}.
\begin{itemize}
\item In order to get \lstinline{rect}, we build a virtual induction proof on \lstinline{T} and analyze that proof term. For example, one can prove ``\lstinline{forall l: natlist, Type}'' by the following tactic:
\begin{lstlisting}
  intro l; induction l; exact bool.
\end{lstlisting}
This tactic will generate the following proof term:
\begin{lstlisting}
  fun l : natlist => natlist_rect (fun _ => Type) bool (fun _ _ _ => bool) l
\end{lstlisting}
Then our tactic analyzes this proof term to get \lstinline{natlist_rect}.
\item In order to get \lstinline{constrs}, we get \lstinline{rect} first and analyze the syntax of its type.
For example,
\begin{lstlisting}
  natlist_rect:
      forall P : natlist -> Type,
        (forall (n0 : nat) (l1 : natlist), P l1 -> P (Cons n0 l1)) ->
        (P Nil) ->
        (forall l : natlist, P l)
\end{lstlisting}
From its assumptions, our tactic can generate
\begin{lstlisting}
  [ _[ [Some nat; None], Cons ]_; _[ [], Nil ]_ ].
\end{lstlisting}
\item In order to get \lstinline{rect_correctness}, we only need to unfold the definitions of \lstinline{rect_correct} and \lstinline{apply_rect}, and prove the conclusion by reflexivity.
\end{itemize}

%% file: formalize.tex
\section{The countability theorem} \label{sec:form}

As mentioned in Section \ref{sec:intro}, the main proof idea is to define a \lstinline{rank} function from inductive type \lstinline{T} to \lstinline{nat}, and prove that $T_n \triangleq \{x: T \mid \mathsf{rank}(x) < n\}$ is countable for any $n$. This conclusion can be proved by induction on $n$. Its induction step is to construct an injection from $T_{n+1}$ to the union of different products of $T_n$, and the latter one is countable since $T_n$ is countable by the induction hypothesis. Using \lstinline{natlist} and \lstinline{bintree} as examples, we can construct injections\footnote{Here we choose not to present the correct \lstinline{rank} function, for reasons of conciseness. The real general \lstinline{rank} function takes more arguments to be specialized on \lstinline{natlist} and \lstinline{bintree}; see Section \ref{sec:form1}. Also, redundant ``\lstinline{* unit}'' and ``\lstinline{+ void}'' are introduced by a uniform recursive definition for convenience.
}:
\begin{lstlisting}[mathescape]
    natlist$_\mathsf{n+1}$ -> 
      nat * (natlist$_\mathsf{n}$ * unit) + (unit + void)
    bintree$_\mathsf{n+1}$ -> 
      nat * (bintree$_\mathsf{n}$ * (bintree$_\mathsf{n}$ * unit)) + (unit + void)
\end{lstlisting}
Using \lstinline{natlist} as an example, this construction of injection takes three steps:
\begin{itemize}
\item Defining a function from \lstinline{natlist} to \lstinline{nat * (natlist * unit) + (unit + void)}:
\begin{lstlisting}
    pattern_match (l: natlist) :=
      match l with
      | Cons n0 l1 => inl (n0, (l1, tt))
                              (* inl chooses the left branch of sum type *)
                              (* tt is the only element of unit *)
      | Nil => inr (inl (tt)) 
                             (* inr chooses the right branch of sum type *)
      end.
\end{lstlisting}
\item Well-definedness of \lstinline{pattern_match}: \\
We thus prove that if \lstinline{rank l < S n} and  \lstinline{pattern_match l = inl (n0, (l1, tt))}, then \lstinline{rank l1 < n}.
Thus, we can define a dependently typed function of the type below based on \lstinline{pattern_match}.
\begin{lstlisting}
    {l: natlist | rank l < S n} -> 
      nat * ({l: natlist | rank l < n} * unit) + (unit + void)
\end{lstlisting}
In other words, we define
\begin{lstlisting}[mathescape]
    pattern_match_DT:
      natlist$_\mathsf{n+1}$ -> nat * (natlist$_\mathsf{n}$ * unit) + (unit + void)
\end{lstlisting}
\item Injective property: \\
We prove that the \lstinline{pattern_match} function we defined is an injection. 
\end{itemize}

In Section \ref{sec:form1}, we introduce our general definition of \lstinline{rank} and \lstinline{pattern_match}.
In Section \ref{sec:form2} and \ref{sec:form3}, we establish the injective property above. Specifically, we first prove that \lstinline{pattern_match} itself is an injection (see Section \ref{sec:form2}) and use that conclusion to prove our final dependently typed version injective (see Section \ref{sec:form3}).
Finally, we summarize our main theorem in Section \ref{sec:mainthm}.

\input{form1}
\input{form2}

\input{form3}
\input{maintheorem}

%% file: form1.tex
\subsection{Definitions} \label{sec:form1}

We first define the function that calculates the union of different products named \lstinline{normtype}.
As shown in the beginning of section \ref{sec:form}, we want to project $T_{n+1}$ into the union of different products of $T_n$.
Also, the non-dependent type version \lstinline{pattern_match} is a function from $T$ to the union of different products of $T$.
Thus, our definition of \lstinline{normtype} is polymorphic.
For example,
\begin{lstlisting}
    normtype natlist [ _[ [Some nat; None], Cons ]_; _[ [], Nil ]_ ] X
      = nat * (X * unit) + (unit + void)
    normtype bintree [ _[ [Some nat; None; None], Node ]_; _[ [], Leaf ]_ ] X
      = nat * (X * (X * unit) + (unit + void)
\end{lstlisting}
In our definition of \lstinline{normtype}, we analyze the inductive definition \lstinline{constrs} \linebreak (which means \lstinline {[ _[ [Some nat; None], cons ]_; _[ [], nil ]_ ]} for \lstinline{natlist}), use product types to represent each branch, and use sum types to connect them. 
For each branch, we use $A$ when we meet \lstinline{Some A} and use $X$ when we meet \lstinline{None}. We can therefore generate all \lstinline{normtype} types with the same syntax tree as $T$ like $T_{n+1}$.

As mentioned in Section \ref{sec:pre}, we need to define all recursive function (e.g. \lstinline{rank}) and pattern match expressions (e.g. \lstinline{pattern_match}) based on the general recursive function.
Specifically, suppose \lstinline{rect} is the general recursive function of type \lstinline{T} with inductive definition \lstinline{constrs}, we define \lstinline{rank} and \lstinline{pattern_match} by filling \lstinline{rect}'s arguments through \lstinline{apply_rect}.

When defining \lstinline{rank}, each argument of \lstinline{rect} is to add recursive calls' results together. For example, \lstinline{size} (see Section \ref{sec:intro}) is the \lstinline{rank} function of \lstinline{bintree}.
We can define it by filling \lstinline{bintree}'s arguments:
\begin{lstlisting}
    size t = bintree_rect _ (fun n0 t1 r1 t2 r2 => r1 + r2 + 1) (1)
\end{lstlisting}
Here, in the recursive branch for constructor ``\lstinline{Node}'', \lstinline{n0} is the label, \lstinline{t1} and \lstinline{t2} are the left and right subtrees respectively, and \lstinline{r1} and \lstinline{r2} are the results of recursive calls: \lstinline{size t1} and \lstinline{size t2}; and in the recursive branch of constructor ``\lstinline{Leaf}'', the return value is a constant \lstinline{1}.
In general, we can define these arguments of \lstinline{rect} based on the syntax of inductive definitions.
Using the example above, the typing information of ``\lstinline{Node}'' is described by \lstinline{[Some nat; None; None]} since \lstinline{Node} has type:
\begin{lstlisting}
    nat -> bintree -> bintree -> bintree
\end{lstlisting}
The first element \lstinline{Some nat} corresponds to \lstinline{n0} above; the second element \lstinline{None} corresponds to \lstinline{t1} and \lstinline{r1} above; and the element  \lstinline{None} corresponds to \lstinline{t2} and \lstinline{r2}.

Defining \lstinline{pattern_match} is more complicated.
Here is how we define the \lstinline{pattern_match} function for \lstinline{natlist} (see the beginning of Section \ref{sec:form}) by filling \lstinline{rect}'s arguments.
\begin{lstlisting}
    pattern_match t =
      natlist_rect _ (fun n0 l1 r1 => inl (n0, (l1, tt))) (inr (inl tt))
\end{lstlisting}
Here, we put \lstinline{inl} in the \lstinline{Cons} branch since it is the first branch, and we put \lstinline{inr} $\circ$ \lstinline{inl} in the \lstinline{Nil} branch since it is the second branch.
That means we cannot define these two arguments of \lstinline{natlist_rect} based only on \lstinline{Cons}'s and \lstinline{Nil}'s typing information.
Specifically, if \lstinline{constrs} can be decomposed into: \lstinline{constrs1 ++ _[arg2,constr2]_ :: constrs3}, then \lstinline{rect}'s arguments for the \lstinline{arg2}-\lstinline{constr2}-branch also depend on the length of \lstinline{constrs1} and \lstinline{constrs3}.
For this reason, the definition of \lstinline{apply_rect} must allow \lstinline{para} (the parameterized way of filling \lstinline{rect}'s arguments) to take \lstinline{constrs1} and \lstinline{constrs3} as its arguments.

In our real definition, \lstinline{apply_rect}'s type is:
\begin{lstlisting}
    apply_rect T constrs P para rect_P : forall (t: T), P constrs t.
\end{lstlisting}
where (we omit \lstinline{para}'s type first and introduce it later)
\begin{lstlisting}
    P: Constrs_type -> T -> Type;
    rect_P: rect_type T constrs (P constrs)
\end{lstlisting}
Here \lstinline{P} defines the dependently typed result type and \lstinline{rect_P} is a specialized \lstinline{rect} for computing results of ``type'' \lstinline{P}.
The most important parameter of \lstinline{apply_rect} is \lstinline{para}. Its type is:
\begin{lstlisting}
    para: forall constrs1 arg2 constr2 constrs3,
      let constrs := rev_append constrs1 (_[arg2, constr2]_ :: constrs3) in
      rect_clause_type arg2 T (P constrs) constr2
\end{lstlisting}
That is: if \lstinline{constrs}, all branches of inductive definitions, can be decomposed into \lstinline{(rev constrs1)}, \lstinline{_[arg2, constr2]_}, and \lstinline{constrs3},
then \lstinline{para} computes \lstinline{rect_P}'s argument for the branch of \lstinline{arg2} and \lstinline{constr2}.
Here, the function \lstinline{rev_append} is the auxiliary function for defining a tail-recursive list reverse provided by Coq's standard library:
\begin{lstlisting}
    rev_append {A} (l l' : list A) : list A :=
      match l with
      | [] => l'
      | a :: l0 => rev_append l0 (a :: l')
      end.
\end{lstlisting}
It ensures that for any \lstinline{l} and \lstinline{l'}, \lstinline{rev_append l l' = rev l ++ l}.
It is critical for us to use
\begin{lstlisting}
    rev_append constrs1 (_[arg2, constr2]_ :: constrs3)
\end{lstlisting}
instead of
\begin{lstlisting}
    constrs1 ++ _[arg2, constr2]_ :: constrs3
\end{lstlisting}
because it is much more convenient for building dependently typed inductive proofs and dependently typed recursive functions.
When we apply an inductive proof on the structure of \lstinline{constrs},
we can easily transform \newline
\lstinline{           rev_append constrs1 (_[arg2, constr2]_ :: constrs3)} \newline
\lstinline{         } to \lstinline{rev_append (_[arg2, constr2]_ :: constrs1) constrs3} \newline
because they are $\beta\eta\iota$-reduction to each other, and so can pass Coq's unification checking.
But it is not the case for \newline
\lstinline{          constrs1 ++ (_[arg2, constr2]_ :: constrs3)}\newline
\lstinline{        } and \lstinline{(constrs1 ++ [ _[arg2, constr2]_ ]) ++ constrs3} \newline
which are equal, and can be proved equal, but they cannot pass Coq’s unification checking.

In the end, we define \lstinline{para}s for \lstinline{rank} and \lstinline{pattern_match}, and define these two functions for generic inductive types based on corresponding \lstinline{para}s and \lstinline{apply_rect}.

%% file: form2.tex
\subsection{Injective property: the simple typed version} \label{sec:form2}

We have defined \lstinline{pattern_match}, a function from \lstinline{T} to \lstinline{normtype T constrs T} (which is a union type of different product types).
We prove it to be an injection in this subsection.
Later, we will use this simplified conclusion to establish our ultimate goal: a \lstinline{Setoid_injection} from \lstinline{T}$_{n + 1}$ to \lstinline{normtype T} \lstinline{constrs T}$_n$.
 
To prove \lstinline{pattern_match} to be injective (formalized as \lstinline{PM_inj} below), we need to perform case analysis over both $a$ and $b$.
\begin{lstlisting} 
    Definition PM_inj T constrs rect: Prop := 
      forall a b: T, pattern_match a = pattern_match b -> a = b.
\end{lstlisting}  
Again, case analysis proofs is nontrivial since we are now reasoning about a generic inductive type, not a concrete one.
We use a specialized \lstinline{rect} to solve the problem:
\begin{lstlisting}
    Definition rect_PM_inj:
      rect_type T constrs
        (fun a => forall b, pattern_match a = pattern_match b -> a = b) 
    :=  rect _.
\end{lstlisting}
Here, \lstinline{rect_PM_inj} is a proof of a big implication proposition, whose conclusion is exactly \lstinline{PM_inj}
and whose assumptions are those case analysis branches.
Thus, the proof of the injective property can be built by filling all those arguments of \lstinline{rect_PM_inj}.
Specifically, we decompose \lstinline{constrs} into a form of \lstinline{rev_append constrs1 constrs2} (at first \lstinline{constrs1 = []} and \lstinline{constrs2 = constrs}) and fill those arguments by an induction over \lstinline{constrs2}.
We successfully avoid dependently typed type-casting since we use \lstinline{rev_append} in this proof and in \lstinline{apply_rect}'s types.
We omit proof details here.

We now finish the definition of a ``simple'' typed function, \lstinline{pattern_match}, and the proof of its injective property. During the proof process, we encounter some problems regarding dependent types and solve them with the help of \lstinline{rev_append}. To construct an injection from $T_{n+1}$ (defined as $\lbrace$\lstinline{x : T |} \lstinline{rank x < S n}$\rbrace$) to \lstinline{normtype T constrs T}$_n$, we need to prove linear arithmetic properties about \lstinline{rank}.
We could prove that dependently typed injective property in a similar way, but the dependent types with irrelevant proofs will trouble us much more. So we choose a different proof strategy.

%% file: form3.tex
\subsection{Injective property: the dependent type version} \label{sec:form3}
\begin{figure}[h]
	\centering
	\includegraphics[width=0.8\textwidth]{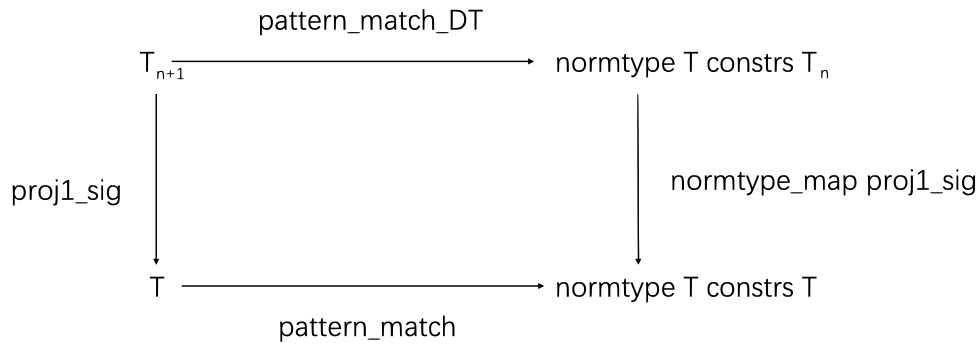}
	\caption{Proof strategy}
	\label{fig}	
\end{figure}

Instead of developing a similar (but more complicated) proof of the injective property for the dependently typed version of \lstinline{pattern_match},
we choose to \emph{use} our injective proof above to build our dependently typed injective proof. In order to prove \lstinline{pattern_match_DT} (the dependently typed version of \lstinline{pattern_match}, a mapping from \lstinline{T}$_{n+1}$ to \lstinline{normtype T constrs T}$_n$) to be injective, it is sufficient to show (Fig. \ref{fig} illustrates this proof strategy):
\begin{itemize}
\item \lstinline{pattern_match} $\circ$ \lstinline{proj1_sig = (normtype_map proj1_sig)} $\circ$ \lstinline{pattern_match_DT};
\item \lstinline{pattern_match} is injective;
\item \lstinline{proj1_sig} is injective.
\end{itemize}

\noindent Here, \lstinline{proj1_sig} (defined by Coq standard library) removes the proof part of Coq's sigma types:
\begin{lstlisting}
    proj1_sig {A} {P: A -> Prop} (e : sig P): A :=
      match e with exist _ a _ => a end
\end{lstlisting}
and we define \lstinline{normtype_map} to apply \lstinline{proj1_sig} on each part of \lstinline{normtype T constrs T}$_n$.

The reasoning behind this proof strategy is straightforward.
By the second and the third condition (\lstinline{pattern_match} and \lstinline{proj1_sig} are injective), we know that  \lstinline{pattern_match} $\circ$ \lstinline{proj1_sig} is injective using theorem \lstinline{injective_compose} (see Section \ref{subsec:count}).
Thus, \lstinline{(normtype_map proj1_sig)} $\circ$ \lstinline{pattern_match_DT} is also injective according to the first condition (the diagram in Fig. \ref{fig} commutes).
In the end, \lstinline{pattern_match_DT} must be an injection due to theorem \lstinline{injective_compose_rev} (see Section \ref{subsec:count}).

We proved \lstinline{pattern_match} to be injective in Section \ref{sec:form3} and \lstinline{proj1_sig} is obviously an injection; thus we only need to prove
\lstinline{pattern_match} $\circ$ \lstinline{proj1_sig = (normtype_map proj1_sig)} $\circ$ \lstinline{pattern_match_DT}, based on our definition of \lstinline{pattern_match_DT} and \lstinline{normtype_map}.

In our Coq formalization, the real type of \lstinline{pattern_match_DT} is:
\begin{lstlisting}[mathescape]
    Definition pattern_match_DT n:
      forall t:T, rank t < S n -> normtype T constrs T$_\mathsf{n}$.
\end{lstlisting}
given type \lstinline{T}, its inductive definition \lstinline{constrs} and its general recursive function \lstinline{rect} and \lstinline{rect_correctness}.
Using \lstinline{natlist} as an example, 
\begin{lstlisting}[mathescape]
    pattern_match_DT n (l: natlist):
       length l < S n -> nat * (natlist$_\mathsf{n}$ * unit) + (unit + void)
    := match l with
        | Cons n0 l1 =>
             fun H: length (Cons n0 l1) < S n =>
                inl (n0, (exist _ l1 SomeProof, tt))
        | Nil =>
             fun H: length Nil < S n =>
                inr (inl (tt)) 
        end.
\end{lstlisting}
The function above may be hard to read. It is equivalent to the following one, which is written in a less dependently typed way.
\begin{lstlisting}
    pattern_match_DT_demo n (l: natlist) (H: length l < S n) :=
      match l with
      | Cons n0 l1 => inl (n0, (exist _ l1 SomeProof, tt))
      | Nil => inr (inl (tt)) 
      end.
\end{lstlisting}
In their \lstinline{Cons} branches, we need to provide a proof of \lstinline{length l1 < n} at the place of \lstinline{SomeProof} given \lstinline{H: length l < S n}.

In order to define such a \lstinline{pattern_match_DT} function for a generic inductive type \lstinline{T}, 
we build its definition based on \lstinline{apply_rect} and a ``\lstinline{para}'' of the following type:
\begin{lstlisting}[mathescape]
    Definition pattern_match_para_DT n T constrs1 arg2 constr2 constrs3:
        let constrs := rev_append constrs1 (_[arg2, constr2]_ :: constrs3) in
        rect_clause_type arg2 T
           (fun t => rank t < S n -> normtype T constrs T$_\mathsf{n}$)
           constr2.
\end{lstlisting}
That is, we define \lstinline{pattern_match_DT} by filling arguments of \lstinline{T}'s general recursive function \lstinline{rect}.
Given that \lstinline{constrs} is decomposed into \lstinline{rev_append constrs1 (_[arg2, constr2]_ :: constrs3)},
this \lstinline{para} above computes the \lstinline{arg2}-{constr2}-argument of \lstinline{rect}.
Again, using the \lstinline{Cons} argument of \lstinline{natlist_rect} as an example, this \lstinline{para} should have the following type:
\begin{lstlisting}[mathescape]
  forall n0 l1,
    length (Cons n0 l1) < S n ->
    normtype natlist constrs_natlist natlist$_\mathsf{n}$
\end{lstlisting}
In the definition of \lstinline{pattern_match_para_DT} and the proof of \lstinline{pattern_match} $\circ$ \lstinline{proj1_sig =} \linebreak \lstinline{(normtype_map proj1_sig)} $\circ$ \lstinline{pattern_match_DT}, we repeatedly use \lstinline{rect_correctness} to reason about \lstinline{rank}. For the sake of space, we omit details here.

%% file: maintheorem.tex
\subsection{Main theorem} \label{sec:mainthm}
We have constructed the injection from \lstinline{T}$_{n+1}$ to \lstinline{normtype T constrs T}$_n$ in Coq. The definition of \lstinline{apply_rect} helps us define most of the functions we need, and most of the intermediate lemmas can be reduced to properties of \lstinline{apply_rect} and the corresponding \lstinline{para}, which greatly simplifies our proof. We use \lstinline{rev_append} to avoid type casting.

In summary, we use \lstinline{SetoidCountable} as an auxiliary definition to prove the countability theorem. Specifically, we prove \lstinline{SetoidCountable T}$_n$ by induction over $n$ and by the injection we just constructed. In the end, our main theorem is:
\begin{lstlisting}
    Variable (T: Type).
    Variable (constrs: Constrs_type).
    Variable (rect: forall P, rect_type T constrs P). 
    Variable (rect_correctness : rect_correct T constrs rect).
    Hypothesis base_countable : 
      Forall_type (fun s => (Forall_type option_Countable) (projT1 s)) constrs.
    Theorem Countable_T : Countable T.
\end{lstlisting}
where the function \lstinline{Forall_type} (provided by Coq's standard library) defines the universal predicates over lists
\begin{lstlisting}
    Inductive Forall_type {A : Type} (P : A -> Type) : list A -> Type :=
        | Forall_type_nil : Forall_type P nil
        | Forall_type_cons : forall (x : A) (l : list A),
                  P x -> Forall_type P l -> Forall_type P (x :: l).
\end{lstlisting}
and the hypothesis \lstinline{base_countable} says that all base types used in the inductive definition are countable.
For example, this theorem proves that \lstinline{natlist} is countable, as long as \lstinline{nat} is countable.

%% file: automation.tex
\section{Automatic proof systems} \label{sec:autom}
To make our result more practical and accessible, we developed Coq tactics to automatically prove inductive types countable.
We define three tactics \lstinline{gen_constrs} , \lstinline{gen_rect} , \lstinline{apply_rect_correctness_gen} to get \lstinline{constrs}, \lstinline{rect} and \lstinline{rect_correctness} from \lstinline{T} (previously mentioned in Section \ref{sec:pre}, see Coq development for more details).
According to the list of conditions and hypothesis, we need in proof (as shown above in Section \ref{sec:mainthm}), the last thing that needs to be done is to prove \lstinline{base_countable}, which means that all base types used in the definition of \lstinline{T} are countable.
This is done using a library of proved countability results and Coq assumptions. Here are two typical applications of our tactic \lstinline{Countable_solver}:
\begin{lstlisting}
    Theorem Countable_expr : Countable expr.
    Proof. intros. Countable_solver. Qed.
    Theorem Countable_list : forall A: Type, Countable A -> Countable (list A).
    Proof. intros. Countable_solver. Qed.
\end{lstlisting}

%% file: relatedwork.tex
\section{Related work} \label{sec:related}
We discuss three related studies in proving relevant properties for inductive types in Coq and one in Isabelle/HOL. What sets our formalization apart from this work is our proof at Coq’s object logic level; only  Deriving \cite{coqPL2020} develops object logic proofs like us. As a result, our proofs and automation instructions do not need other additional axioms, libraries, or tools, and can run with high efficiency.  
\begin{itemize}
    \item Theory Countable\cite{HOLlibrary} : 
    These researchers also used injection to represent the \lstinline{Countable} relation. Their main idea was to construct an injection from data types to old data types which are countable. They could automatically prove the countability of data types which had nested and mutual recursion, and used other data types. However, the process and automatic tactics were formalized in a meta language.
    \item Autosubst\cite{schafer2015autosubst} \& Autosubst2\cite{stark2019autosubst} : 
    Autosubst can automatically generate the substitution operations for a custom inductive type of terms, and prove the corresponding substitution lemmas. The library gives the enumerability of De Brujin substitution algebra\cite{schafer2015completeness}. Autosubst offers tactics that implement the normalization and decision procedure. They believe that it is hard to maintain or extend Ltac code, so that they proposed a new implementation of Autosubst which comes in the form of a code generator to generate Coq proof terms, and at the same time extends Autosubst’s input language to mutual inductive sorts with multiple sorts of variables. Autosubst and Autosubst2 develop formalized proofs as tactics and external proof term generators, respectively, which can be treated as meta-level mappings from the syntax of inductive definitions to countability proof terms. In comparison, our proof is a Coq object-level mapping from inductive definitions to countability, but we do not support mutually inductive types at present.
    \item Undecidability\cite{forster2019synthetic} : 
    Forster et. al. formalized the computational undecidability of the validity, satisfiability, and provability of first-order formulas following a synthetic approach based on the computation native to Coq’s constructive type theory. They extended the library in 2020, to present a comprehensive analysis of the computational content of completeness theorems for first-order logic, considering various semantics and deduction systems\cite{forster2020completeness}. They proved first-order logic’s propositions countable in their completeness proof. They also formalized the syntax of inductive definitions\footnote{This general formalization appears in their Coq development but they do not mention that in their paper.}, as we did, but they do not provide a general theorem of countability.
    \item Deriving \cite{coqPL2020} :
    Deriving proved inductive types countable in Coq.
    Deriving uses \lstinline{countType} in the MathComp library, which provides a different definition of injection and \lstinline{Countable}.
    Although this alternative definition requires injections to be Coq-computable functions, it is not a significant drawback comparing with our definitions when applying inductive type's countability.
    Deriving supports the proof of countability for mutually inductive types and nested inductive types.
    Their proof strategy is different from ours: they built an injection from every inductive type $T$ to \emph{finite-width trees}, which they defined as a Coq type and proved countable.
    More significantly, they provided two formalization of inductive definitions. One is like ours:
     \begin{lstlisting}
     constrs: list (sigT (fun arg => constr_type arg T));
     \end{lstlisting}
the other uses indices to number the constructors.
In other words, the latter formalization is a deep embedding of the meta language (the syntax of inductive types) into Coq's object language.
They carefully used computable functions over the deeply embedded meta language (since natural numbers' equality tests are computable, but types' equality tests are not computable) in their definitions and used the connection between these two formalizations to compute their final proof term.
In their proofs, they need to reason about indices' equalities, types' equalities and relevant dependent type issues---if two Coq types $T_1 = T_2$, $T_1$'s elements are not automatically recognized as $T_2$'s elements by Coq's type checker.
    In comparison, our work shows that inductions over constructor lists do prove the conclusion using the ``\lstinline{rev-append} trick'', and we do not need number-indexing and heavy-weighted proof reflection to bypass related difficulties in dependently type proofs.
\end{itemize}

%% file: conclusion.tex
\section{Conclusions} \label{sec:concl}

We proved in Coq that a first-order inductive type is countable as long as all base types used in the definition are countable. Our definitions and proofs are all axiom-free. We provide an alternative way of thinking about solving dependent types at the object level. We developed very efficient tactics which use this countability theorem to prove concrete inductive types to be countable. Our formalization and automatic tactic still have room for expansion in the future. For example, our tactics do not work when applied to mutually recursive types. We believe that it is plausible to transform the mutually recursive types into primitive recursive types \cite{113772} and enhance our tactics.
Our Coq development can be found at:

\ \ \ \ \ \ \ \ \ \ \ \ \ \ \ \ \ \ \ \url{https://github.com/QinxiangCao/Countable_PaperSubmission}

%% file: paper.bbl
\begin{thebibliography}{10}
\providecommand{\bibitemdeclare}[2]{}
\providecommand{\surnamestart}{}
\providecommand{\surnameend}{}
\providecommand{\urlprefix}{Available at }
\providecommand{\url}[1]{\texttt{#1}}
\providecommand{\href}[2]{\texttt{#2}}
\providecommand{\urlalt}[2]{\href{#1}{#2}}
\providecommand{\doi}[1]{doi:\urlalt{http://dx.doi.org/#1}{#1}}
\providecommand{\bibinfo}[2]{#2}

\bibitemdeclare{misc}{HOLlibrary}
\bibitem{HOLlibrary}
\bibinfo{author}{Jasmin~Blanchette \surnamestart Alexander~Krauss\surnameend,
  Brian~Huffman}: \emph{\bibinfo{title}{This is a library of Countable Theory
  in {Isabelle}/{HOL}}}.
\newblock
  \bibinfo{howpublished}{\url{https://devel.isa-afp.org/browser_info/current/HOL/HOL-Library/Countable.html}}.

\bibitemdeclare{article}{coqPL2020}
\bibitem{coqPL2020}
\bibinfo{author}{A.~A. \surnamestart de~Amorim\surnameend}
  (\bibinfo{year}{2020}): \emph{\bibinfo{title}{Deriving Instances with
  Dependent Types}}.
\newblock {\sl \bibinfo{journal}{CoqPL 2020}}.

\bibitemdeclare{inproceedings}{anand2014generic}
\bibitem{anand2014generic}
\bibinfo{author}{Abhishek \surnamestart Anand\surnameend} \&
  \bibinfo{author}{Vincent \surnamestart Rahli\surnameend}
  (\bibinfo{year}{2014}): \emph{\bibinfo{title}{A Generic Approach to Proofs
  about Substitution}}.
\newblock In \bibinfo{editor}{Amy~P. \surnamestart Felty\surnameend} \&
  \bibinfo{editor}{Brigitte \surnamestart Pientka\surnameend}, editors: {\sl
  \bibinfo{booktitle}{Proceedings of the 2014 International Workshop on Logical
  Frameworks and Meta-languages: Theory and Practice, {LFMTP} '14, Vienna,
  Austria, July 17, 2014}}, \bibinfo{publisher}{{ACM}}, pp. \bibinfo{pages}{5:
  1--5: 8}, \doi{10.1145/2631172.2631177}.

\bibitemdeclare{article}{aydemir2010lngen}
\bibitem{aydemir2010lngen}
\bibinfo{author}{Brian \surnamestart Aydemir\surnameend} \&
  \bibinfo{author}{Stephanie \surnamestart Weirich\surnameend}
  (\bibinfo{year}{2010}): \emph{\bibinfo{title}{LNgen: Tool support for locally
  nameless representations}}.

\bibitemdeclare{article}{barras1999coq}
\bibitem{barras1999coq}
\bibinfo{author}{Bruno \surnamestart Barras\surnameend},
  \bibinfo{author}{Samuel \surnamestart Boutin\surnameend},
  \bibinfo{author}{Cristina \surnamestart Cornes\surnameend},
  \bibinfo{author}{Judica{\"e}l \surnamestart Courant\surnameend},
  \bibinfo{author}{Yann \surnamestart Coscoy\surnameend},
  \bibinfo{author}{David \surnamestart Delahaye\surnameend},
  \bibinfo{author}{Daniel \surnamestart de~Rauglaudre\surnameend},
  \bibinfo{author}{Jean-Christophe \surnamestart Filli{\^a}tre\surnameend},
  \bibinfo{author}{Eduardo \surnamestart Gim{\'e}nez\surnameend},
  \bibinfo{author}{Hugo \surnamestart Herbelin\surnameend} et~al.
  (\bibinfo{year}{1999}): \emph{\bibinfo{title}{The {Coq} proof assistant
  reference manual}}.
\newblock {\sl \bibinfo{journal}{INRIA, version}}
  \bibinfo{volume}{6}(\bibinfo{number}{11}).

\bibitemdeclare{inproceedings}{cao2017bringing}
\bibitem{cao2017bringing}
\bibinfo{author}{Qinxiang \surnamestart Cao\surnameend},
  \bibinfo{author}{Santiago \surnamestart Cuellar\surnameend} \&
  \bibinfo{author}{Andrew~W. \surnamestart Appel\surnameend}
  (\bibinfo{year}{2017}): \emph{\bibinfo{title}{Bringing Order to the
  Separation Logic Jungle}}.
\newblock In \bibinfo{editor}{Bor{-}Yuh~Evan \surnamestart Chang\surnameend},
  editor: {\sl \bibinfo{booktitle}{Programming Languages and Systems - 15th
  Asian Symposium, {APLAS} 2017, Suzhou, China, November 27-29, 2017,
  Proceedings}}, {\sl \bibinfo{series}{Lecture Notes in Computer Science}}
  \bibinfo{volume}{10695}, \bibinfo{publisher}{Springer}, pp.
  \bibinfo{pages}{190--211}, \doi{10.1007/978-3-319-71237-6\_10}.

\bibitemdeclare{inproceedings}{chlipala2008parametric}
\bibitem{chlipala2008parametric}
\bibinfo{author}{Adam \surnamestart Chlipala\surnameend}
  (\bibinfo{year}{2008}): \emph{\bibinfo{title}{Parametric higher-order
  abstract syntax for mechanized semantics}}.
\newblock In \bibinfo{editor}{James \surnamestart Hook\surnameend} \&
  \bibinfo{editor}{Peter \surnamestart Thiemann\surnameend}, editors: {\sl
  \bibinfo{booktitle}{Proceeding of the 13th {ACM} {SIGPLAN} international
  conference on Functional programming, {ICFP} 2008, Victoria, BC, Canada,
  September 20-28, 2008}}, \bibinfo{publisher}{{ACM}}, pp.
  \bibinfo{pages}{143--156}, \doi{10.1145/1411204.1411226}.

\bibitemdeclare{book}{ebbinghaus2013mathematical}
\bibitem{ebbinghaus2013mathematical}
\bibinfo{author}{Heinz{-}Dieter \surnamestart Ebbinghaus\surnameend},
  \bibinfo{author}{J{\"{o}}rg \surnamestart Flum\surnameend} \&
  \bibinfo{author}{Wolfgang \surnamestart Thomas\surnameend}
  (\bibinfo{year}{1994}): \emph{\bibinfo{title}{Mathematical logic {(2.}
  ed.)}}.
\newblock \bibinfo{series}{Undergraduate texts in mathematics},
  \bibinfo{publisher}{Springer}, \doi{10.1007/978-1-4757-2355-7}.

\bibitemdeclare{inproceedings}{forster2019synthetic}
\bibitem{forster2019synthetic}
\bibinfo{author}{Yannick \surnamestart Forster\surnameend},
  \bibinfo{author}{Dominik \surnamestart Kirst\surnameend} \&
  \bibinfo{author}{Gert \surnamestart Smolka\surnameend}
  (\bibinfo{year}{2019}): \emph{\bibinfo{title}{On synthetic undecidability in
  coq, with an application to the entscheidungsproblem}}.
\newblock In \bibinfo{editor}{Assia \surnamestart Mahboubi\surnameend} \&
  \bibinfo{editor}{Magnus~O. \surnamestart Myreen\surnameend}, editors: {\sl
  \bibinfo{booktitle}{Proceedings of the 8th {ACM} {SIGPLAN} International
  Conference on Certified Programs and Proofs, {CPP} 2019, Cascais, Portugal,
  January 14-15, 2019}}, \bibinfo{publisher}{{ACM}}, pp.
  \bibinfo{pages}{38--51}, \doi{10.1145/3293880.3294091}.

\bibitemdeclare{inproceedings}{forster2020completeness}
\bibitem{forster2020completeness}
\bibinfo{author}{Yannick \surnamestart Forster\surnameend},
  \bibinfo{author}{Dominik \surnamestart Kirst\surnameend} \&
  \bibinfo{author}{Dominik \surnamestart Wehr\surnameend}
  (\bibinfo{year}{2020}): \emph{\bibinfo{title}{Completeness Theorems for
  First-Order Logic Analysed in Constructive Type Theory}}.
\newblock In \bibinfo{editor}{Sergei~N. \surnamestart Art{\"{e}}mov\surnameend}
  \& \bibinfo{editor}{Anil \surnamestart Nerode\surnameend}, editors: {\sl
  \bibinfo{booktitle}{Logical Foundations of Computer Science - International
  Symposium, {LFCS} 2020, Deerfield Beach, FL, USA, January 4-7, 2020,
  Proceedings}}, {\sl \bibinfo{series}{Lecture Notes in Computer Science}}
  \bibinfo{volume}{11972}, \bibinfo{publisher}{Springer}, pp.
  \bibinfo{pages}{47--74}, \doi{10.1007/978-3-030-36755-8\_4}.

\bibitemdeclare{inproceedings}{113772}
\bibitem{113772}
\bibinfo{author}{Peter \surnamestart {Freyd}\surnameend}
  (\bibinfo{year}{1990}): \emph{\bibinfo{title}{Recursive types reduced to
  inductive types}}.
\newblock In: {\sl \bibinfo{booktitle}{[1990] Proceedings. Fifth Annual IEEE
  Symposium on Logic in Computer Science}}, pp. \bibinfo{pages}{498--507},
  \doi{10.1109/LICS.1990.113772}.

\bibitemdeclare{inproceedings}{316071}
\bibitem{316071}
\bibinfo{author}{M.~\surnamestart Hofmann\surnameend} \&
  \bibinfo{author}{T.~\surnamestart Streicher\surnameend}
  (\bibinfo{year}{1994}): \emph{\bibinfo{title}{The groupoid model refutes
  uniqueness of identity proofs}}.
\newblock In: {\sl \bibinfo{booktitle}{Proceedings Ninth Annual IEEE Symposium
  on Logic in Computer Science}}, pp. \bibinfo{pages}{208--212},
  \doi{10.1109/LICS.1994.316071}.

\bibitemdeclare{inproceedings}{lee2012gm}
\bibitem{lee2012gm}
\bibinfo{author}{Gyesik \surnamestart Lee\surnameend},
  \bibinfo{author}{Bruno~C. \surnamestart d.~S.~Oliveira\surnameend},
  \bibinfo{author}{Sungkeun \surnamestart Cho\surnameend} \&
  \bibinfo{author}{Kwangkeun \surnamestart Yi\surnameend}
  (\bibinfo{year}{2012}): \emph{\bibinfo{title}{GMeta: {A} Generic Formal
  Metatheory Framework for First-Order Representations}}.
\newblock In \bibinfo{editor}{Helmut \surnamestart Seidl\surnameend}, editor:
  {\sl \bibinfo{booktitle}{Programming Languages and Systems - 21st European
  Symposium on Programming, {ESOP} 2012, Held as Part of the European Joint
  Conferences on Theory and Practice of Software, {ETAPS} 2012, Tallinn,
  Estonia, March 24 - April 1, 2012. Proceedings}}, {\sl
  \bibinfo{series}{Lecture Notes in Computer Science}} \bibinfo{volume}{7211},
  \bibinfo{publisher}{Springer}, pp. \bibinfo{pages}{436--455},
  \doi{10.1007/978-3-642-28869-2\_22}.

\bibitemdeclare{inproceedings}{DBLP:conf/cade/MouraKADR15}
\bibitem{DBLP:conf/cade/MouraKADR15}
\bibinfo{author}{Leonardo~Mendon{\c{c}}a \surnamestart de~Moura\surnameend},
  \bibinfo{author}{Soonho \surnamestart Kong\surnameend},
  \bibinfo{author}{Jeremy \surnamestart Avigad\surnameend},
  \bibinfo{author}{Floris \surnamestart van Doorn\surnameend} \&
  \bibinfo{author}{Jakob \surnamestart von Raumer\surnameend}
  (\bibinfo{year}{2015}): \emph{\bibinfo{title}{The Lean Theorem Prover (System
  Description)}}.
\newblock In \bibinfo{editor}{Amy~P. \surnamestart Felty\surnameend} \&
  \bibinfo{editor}{Aart \surnamestart Middeldorp\surnameend}, editors: {\sl
  \bibinfo{booktitle}{Automated Deduction - {CADE-25} - 25th International
  Conference on Automated Deduction, Berlin, Germany, August 1-7, 2015,
  Proceedings}}, {\sl \bibinfo{series}{Lecture Notes in Computer Science}}
  \bibinfo{volume}{9195}, \bibinfo{publisher}{Springer}, pp.
  \bibinfo{pages}{378--388}, \doi{10.1007/978-3-319-21401-6\_26}.

\bibitemdeclare{book}{nederpelt2014type}
\bibitem{nederpelt2014type}
\bibinfo{author}{Rob \surnamestart Nederpelt\surnameend} \&
  \bibinfo{author}{Herman \surnamestart Geuvers\surnameend}
  (\bibinfo{year}{2014}): \emph{\bibinfo{title}{Type theory and formal proof:
  an introduction}}.
\newblock \bibinfo{publisher}{Cambridge University Press},
  \doi{10.1017/CBO9781139567725}.

\bibitemdeclare{incollection}{paulin2015introduction}
\bibitem{paulin2015introduction}
\bibinfo{author}{Christine \surnamestart Paulin-Mohring\surnameend}
  (\bibinfo{year}{2015}): \emph{\bibinfo{title}{{Introduction to the Calculus
  of Inductive Constructions}}}.
\newblock In \bibinfo{editor}{Bruno~Woltzenlogel \surnamestart
  Paleo\surnameend} \& \bibinfo{editor}{David \surnamestart
  Delahaye\surnameend}, editors: {\sl \bibinfo{booktitle}{{All about Proofs,
  Proofs for All}}}, {\sl \bibinfo{series}{Studies in Logic (Mathematical logic
  and foundations)}}~\bibinfo{volume}{55}, \bibinfo{publisher}{{College
  Publications}}.
\newblock \urlprefix\url{https://hal.inria.fr/hal-01094195}.

\bibitemdeclare{inproceedings}{polonowski2013automatically}
\bibitem{polonowski2013automatically}
\bibinfo{author}{Emmanuel \surnamestart Polonowski\surnameend}
  (\bibinfo{year}{2013}): \emph{\bibinfo{title}{Automatically Generated
  Infrastructure for De Bruijn Syntaxes}}.
\newblock In \bibinfo{editor}{Sandrine \surnamestart Blazy\surnameend},
  \bibinfo{editor}{Christine \surnamestart Paulin{-}Mohring\surnameend} \&
  \bibinfo{editor}{David \surnamestart Pichardie\surnameend}, editors: {\sl
  \bibinfo{booktitle}{Interactive Theorem Proving - 4th International
  Conference, {ITP} 2013, Rennes, France, July 22-26, 2013. Proceedings}}, {\sl
  \bibinfo{series}{Lecture Notes in Computer Science}} \bibinfo{volume}{7998},
  \bibinfo{publisher}{Springer}, pp. \bibinfo{pages}{402--417},
  \doi{10.1007/978-3-642-39634-2\_29}.

\bibitemdeclare{inproceedings}{schafer2015completeness}
\bibitem{schafer2015completeness}
\bibinfo{author}{Steven \surnamestart Sch{\"{a}}fer\surnameend},
  \bibinfo{author}{Gert \surnamestart Smolka\surnameend} \&
  \bibinfo{author}{Tobias \surnamestart Tebbi\surnameend}
  (\bibinfo{year}{2015}): \emph{\bibinfo{title}{Completeness and Decidability
  of de Bruijn Substitution Algebra in Coq}}.
\newblock In \bibinfo{editor}{Xavier \surnamestart Leroy\surnameend} \&
  \bibinfo{editor}{Alwen \surnamestart Tiu\surnameend}, editors: {\sl
  \bibinfo{booktitle}{Proceedings of the 2015 Conference on Certified Programs
  and Proofs, {CPP} 2015, Mumbai, India, January 15-17, 2015}},
  \bibinfo{publisher}{{ACM}}, pp. \bibinfo{pages}{67--73},
  \doi{10.1145/2676724.2693163}.

\bibitemdeclare{inproceedings}{schafer2015autosubst}
\bibitem{schafer2015autosubst}
\bibinfo{author}{Steven \surnamestart Sch{\"{a}}fer\surnameend},
  \bibinfo{author}{Tobias \surnamestart Tebbi\surnameend} \&
  \bibinfo{author}{Gert \surnamestart Smolka\surnameend}
  (\bibinfo{year}{2015}): \emph{\bibinfo{title}{Autosubst: Reasoning with de
  Bruijn Terms and Parallel Substitutions}}.
\newblock In \bibinfo{editor}{Christian \surnamestart Urban\surnameend} \&
  \bibinfo{editor}{Xingyuan \surnamestart Zhang\surnameend}, editors: {\sl
  \bibinfo{booktitle}{Interactive Theorem Proving - 6th International
  Conference, {ITP} 2015, Nanjing, China, August 24-27, 2015, Proceedings}},
  {\sl \bibinfo{series}{Lecture Notes in Computer Science}}
  \bibinfo{volume}{9236}, \bibinfo{publisher}{Springer}, pp.
  \bibinfo{pages}{359--374}, \doi{10.1007/978-3-319-22102-1\_24}.

\bibitemdeclare{inproceedings}{stark2019autosubst}
\bibitem{stark2019autosubst}
\bibinfo{author}{Kathrin \surnamestart Stark\surnameend},
  \bibinfo{author}{Steven \surnamestart Sch{\"{a}}fer\surnameend} \&
  \bibinfo{author}{Jonas \surnamestart Kaiser\surnameend}
  (\bibinfo{year}{2019}): \emph{\bibinfo{title}{Autosubst 2: reasoning with
  multi-sorted de Bruijn terms and vector substitutions}}.
\newblock In \bibinfo{editor}{Assia \surnamestart Mahboubi\surnameend} \&
  \bibinfo{editor}{Magnus~O. \surnamestart Myreen\surnameend}, editors: {\sl
  \bibinfo{booktitle}{Proceedings of the 8th {ACM} {SIGPLAN} International
  Conference on Certified Programs and Proofs, {CPP} 2019, Cascais, Portugal,
  January 14-15, 2019}}, \bibinfo{publisher}{{ACM}}, pp.
  \bibinfo{pages}{166--180}, \doi{10.1145/3293880.3294101}.

\bibitemdeclare{article}{0Deriving}
\bibitem{0Deriving}
\bibinfo{author}{Christian \surnamestart Sternagel\surnameend} \&
  \bibinfo{author}{Ren{\'{e}} \surnamestart Thiemann\surnameend}
  (\bibinfo{year}{2015}): \emph{\bibinfo{title}{Deriving class instances for
  datatypes}}.
\newblock {\sl \bibinfo{journal}{Arch. Formal Proofs}} \bibinfo{volume}{2015}.
\newblock \urlprefix\url{https://www.isa-afp.org/entries/Deriving.shtml}.

\end{thebibliography}
